\documentclass{ws-procs9x6}

\def\etal {{\it et al.}}

\begin{document}

\title{LORENTZ AND CPT VIOLATION IN THE HYDROGEN SPECTRUM}

\author{G.S.\ ADKINS$^*$}

\address{Department of Physics and Astronomy, Franklin \& Marshall College,\\
Lancaster, PA 17604, USA\\
$^*$E-mail: gadkins@fandm.edu}

\author{T.J.\ YODER}

\address{Department of Physics, Massachusetts Institute of Technology,\\
Cambridge, MA 02139, USA\\
E-mail: tjyoder@mit.edu}

\begin{abstract}
We have studied the effect of hypothetical violations of Lorentz and CPT symmetry by calculating the corrections to the energy levels of hydrogen induced by the Standard-Model Extension (SME).  Hydrogen studies are interesting because the energy levels of hydrogen can be measured with great precision and the theory for hydrogen based on the Standard Model (SM) is well understood.  We obtained corrections through order $\alpha^2$ times the SME parameters for all levels of hydrogen and applied them to determine the SME corrections to the transition frequency for the $2S-1S$ transition.
\end{abstract}

\bodymatter

\section{Introduction}

Hydrogen is an interesting system for the study of beyond-the-SM effects because it is well-understood theoretically,\cite{Eides07} high-precision experiments are available, and it is amenable to the calculation of hypothetical new effects using standard methods.  The leading SME corrections--at $O(\alpha^0)$ times SME parameters--to the hydrogen energies have been known for some time.\cite{Bluhm99}  We have obtained the next order of corrections--at $O(\alpha^2)$ times SME parameters.\cite{Yoder12}  There were several reasons for pursuing this calculation.  One is that the lowest order SME correction depends on only a few combinations of SME coefficients.  Perhaps those combinations are zero or small and the leading order effect occurs at higher order.  Certainly we did find that the higher order result has more structure than the lowest order one in terms of dependence on SME parameters and on the quantum numbers.  In addition, such a calculation is useful for learning more about the theory.  It is difficult to anticipate what issues or problems might arise in a higher order calculation without actually doing the calculation.  Hydrogen is appropriate for the exploration of such issues due to its simplicity.  We also explored the possibility of placing new constraints on the SME coefficients through use of the higher-order results.

There were three main parts in our calculation.  First, we needed to find an effective Hamiltonian to describe the interaction between the constituent fermions in hydrogen with the background SME vectors and tensors.  Next, we needed to evaluate the energy shifts due to the new interactions in appropriate states of the hydrogen atom.  Finally, we needed to calculate the energy shifts coming from SME modifications to the electromagnetic binding potential.  We address these three parts in turn.

We work in the context of standard quantum electrodynamics (QED) extended by the minimal SME.  The appropriate Lagrangian has the form \cite{Lagrangian}
\begin{equation}
{\cal L} = \frac{1}{2} i \bar \psi \Gamma^\nu \overleftrightarrow{D_\nu} \psi - \bar \psi M \psi
\end{equation}
where $D_\nu = \partial_\nu + i q A_\nu$ is the covariant derivative, $F_{\mu \nu} = \partial_\mu A_\nu - \partial_\nu A_\mu$, and $\Gamma^\nu$ and $M$ contain the SME coefficients $a^\mu$, $b^\mu$, $c^{\mu \nu}$, etc.:
\begin{eqnarray}
\Gamma^\nu &=& \gamma^\nu + \Gamma_1^\nu = \gamma^\nu + c^{\mu \nu} \gamma_\mu + d^{\mu \nu} \gamma_5 \gamma_\mu + e^\nu + i f^\nu \gamma_5 + \frac{1}{2} g^{\lambda \mu \nu} \sigma_{\lambda \mu} , \cr
M &=& m + M_1 = m + a^\mu \gamma_\mu + b^\mu \gamma_5 \gamma_\mu + \frac{1}{2} H^{\mu \nu} \sigma_{\mu \nu}.
\end{eqnarray}

The effective fermion Hamiltonian arising from the SME can be written as an expansion in the fermion momentum since the characteristic scale of momentum in a non-relativistic Coulombic bound state is $p \sim m \alpha$.  We will need the expansion up to terms of order $(p/m)^2$.  To that order, the effective Hamiltonian can be obtained by a simple non-relativistic expansion of the expectation value of the self-energy operator $\Sigma(p)$:
\begin{equation} \label{effective}
\overline u_p \Sigma(p) u_p \rightarrow \psi^\dagger H_\mathrm{eff} \psi ,
\end{equation}
where $\Sigma(p) \equiv M_1 - \Gamma_1^\mu p_\mu$ comes from the part of ${\cal L}$ that is bilinear in fermions.  We use the Dirac representation for gamma matrices and the non-relativistic spinor normalization
\begin{equation}
u_p = \left ( \frac{\omega_p+m}{2 \omega_p} \right )^{1/2} \begin{pmatrix} \psi \\ \frac{\vec \sigma \cdot \vec p}{\omega_p+m} \psi \end{pmatrix}
\end{equation}
where $\omega_p=(m^2+\vec p\,^2)^{1/2}$.  It is a simple matter now to expand the matrix elements of various Dirac matrices.  We find, for example, $\overline u_p \sigma^{i j} u_p \approx \epsilon_{i j k} \psi^\dagger \left \{ \sigma^k - \vec \sigma \cdot \vec p p^k/(2 m^2) \right \} \psi$.  We use the non-covariant normalization $u_p^\dagger u_p = 1$ so that $\psi$ has the usual non-relativistic normalization.  For the SME contribution to the effective Hamiltonian we find
\begin{equation}
H_\mathrm{eff} = \left (A+B_k \sigma^k \right ) + \left ( C_i + D_{i k} \sigma^k \right ) \frac{p^i}{m} + \left ( E_{i j} + F_{i j k} \sigma^k \right ) \frac{p^i p^j}{m^2} + \cdots
\end{equation}
where $A=a_0-m e_0-m c_{00}$, $B_k = -b_k + m d_{k 0} - \frac{1}{2} \epsilon_{k a b} \left (m g_{a b 0} - H_{a b} \right )$, $E_{i j} = -m c_{i j} - \frac{1}{2} m c_{0 0} \delta_{i j}$, and $F_{i j k} = \delta_{j k} \tilde d_i + \frac{1}{2} \epsilon_{j k a} \left ( \epsilon_{a i p} b_p - 2m g_{a 0 i} - m g_{a i 0} \right )$ with $\tilde d_i = m d_{0 i} + \frac{1}{2} m d_{i 0} - \frac{1}{4} \epsilon_{i a b} H_{a b}$.  We do not show the $C^i$ and $D^{i k}$ coefficients because they do not contribute to the final result.  The masses and coefficients here are those appropriate to the electron.  There is a similar contribution to $H_\mathrm{eff}$ coming from the proton, but its $O(\alpha^2 \cdot \mathrm{SME})$ contribution is suppressed by a relative factor of the square of the electron to proton mass ratio and is not shown here.  This effective Hamiltonian agrees with that found earlier using the Foldy-Wouthuysen transformation.\cite{FWtransform}

Having obtained the required effective Hamiltonian, we are in a position to calculate SME-induced corrections to hydrogen energy levels.  We couple angular momenta in the usual way: orbital to electron spin first ($\vec L + \vec S_e = \vec J$), then to proton spin ($\vec J + \vec S_p = \vec F$).  States are labeled by quantum numbers $n$, $\ell$, $j$, $f$, and $m_f$: $\vert n, \ell, j, f, m_f \rangle$.  In the presence of the usual perturbations giving rise to fine structure, the Lamb shift, hyperfine splitting, and Zeeman splitting (in the presence of any magnetic field), these states are non-degenerate.  We use standard non-degenerate perturbation theory to calculate the effect of $H_\mathrm{eff}$.

The correction at $O(\alpha^0 \cdot \mathrm{SME})$ is given by $\langle \left ( A+B_k \sigma^k \right ) \rangle$ (we are including electron effects only).  The expectation value of the electron spin matrix is conveniently worked out by use of the Wigner-Ekhart theorem, leading to 
\begin{equation}
\Delta E^0_\mathrm{electron} = A + \xi_e B_3 m_f
\end{equation}
where $\xi_e$ is the reduced matrix element $\xi_e = 4 (j-\ell)(4j-2f+1)/((2\ell+1)(2j+1))$.  (The proton effect is just as easy to calculate but we do not show it here.)  This contribution was first obtained by Bluhm {\it et al.}\cite{Bluhm99}

There is no correction at $O(\alpha^1 \cdot \mathrm{SME})$ because the expectation value $\langle \vec p \, \rangle$ vanishes in states of definite parity.

At $O(\alpha^2 \cdot \mathrm{SME})$ we need the expectation values of $p^i p^j$ and $p^i p^j \sigma^k$.  We found it useful to factor the radial and angular parts: $p^i p^j = p^2 \hat p^i \hat p^j$ and use the fact that the wave functions factorize into radial and angular parts as well.  The radial expectation value was done by way of the virial theorem: $\langle p^2 \rangle_n = -2 m E_n = m^2 \alpha^2/n^2$.  The angular evaluation was facilitated by the separation of $\hat p^i \hat p^j$ into parts having $\ell=2$ and $\ell=0$: $\hat p^i \hat p^j = \left ( \hat p^i \hat p^j - \delta_{i j}/3 \right ) + \delta_{i j}/3$.  After a lengthy exercise in angular momentum algebra we obtained our final result for the electron part of the energy correction at this order:
\begin{eqnarray}
\Delta E^2_\mathrm{electron} &= \Big \{ -\frac{5}{6} \tilde c_{0 0} + \frac{(2j+3)(2j-1)}{16j(j+1)} \left ( \frac{1}{3} - \frac{m_f^2}{f(f+1)} \right ) \tilde c_Q \cr
&+ \frac{m_f}{2f+1} \left ( \frac{2}{2j+1} \tilde d_3 -2(j-\ell) \tilde g_{D3} \right ) \Big \} \left ( \frac{\alpha}{n} \right )^2 
\end{eqnarray}
where $\tilde c_{0 0}=m \delta_{i j} c_{i j}$, $\tilde c_Q= m \delta^Q_{i j} c_{i j}$ with $\delta^Q_{i j} = \delta_{i 1} \delta_{j 1} + \delta_{i 2} \delta_{j 2} - 2 \delta_{i 3} \delta_{j 3}$, and $\tilde g_{D3} = -b_3+m(g_{102}-g_{201}+g_{120})$.

Another energy correction at $O(\alpha^2 \cdot \mathrm{SME})$ comes from the photon sector.  The photon propagator is corrected by SME interactions, which lead to corrections to the potential energy between the electron and proton.  For example, the correction to the potential involving $k_F$ was found by Bailey and Kosteleck\'y:\cite{Bailey04}
$\Delta V = (\alpha/r) (k_F)_{i 0 j 0} \left ( \hat x_i \hat x_j - \delta_{i j} \right )$.  The corresponding energy correction is
\begin{equation}
\Delta E^2_\mathrm{photon} = \Big \{ -\frac{2}{3} \tilde \kappa_0 - \frac{(2j+3)(2j-1)}{16j(j+1)}\left ( \frac{1}{3} - \frac{m_f^2}{f(f+1)} \right ) \tilde \kappa_Q \Big \} \left ( \frac{\alpha}{n} \right )^2 ,
\end{equation}
where $\tilde \kappa_0=m \delta_{i j} (k_F)_{i 0 j 0}$ and $\tilde \kappa_Q = m \delta^Q_{i j} (k_F)_{i 0 j 0}$.

An interesting application of our result is to the 2S-1S transition in hydrogen, which has been measured to extreme precision.\cite{Parthey11}  The SME correction to this energy splitting due to both electron and photon effects is
\begin{equation}
\Delta E(\mathrm{2S-1S})=\Big \{ \frac{5}{2} \tilde c_{0 0} + 2 \tilde \kappa_0 - m_f \left ( \tilde d_3 - \tilde g_{D 3} \right ) \Big \} \frac{\alpha^2}{4} .
\end{equation}
This correction and, of course, experimental input, has been used to set a bound on the diagonal spatial components of the SME tensor $c_{\mu \nu}$.\cite{Altschul10}


\begin{thebibliography}{xx}

\bibitem{Eides07}
M.I.\ Eides, H.\ Grotch, and V.A.\ Shelyuto, {\it Theory of Light Hydrogenic Bound States}, Springer, Berlin, 2007.

\bibitem{Bluhm99} R.\ Bluhm, V.A.\ Kosteleck\'y, and N.\ Russell, Phys.\ Rev.\ Lett.\ {\bf 82}, 2254 (1999).

\bibitem{Yoder12} T.J.\ Yoder and G.S.\ Adkins, Phys.\ Rev.\ D {\bf 86}, 116005 (2012).

\bibitem{Lagrangian} D.\ Colladay and V.\ A.\ Kosteleck\'y, Phys.\ Rev.\ D {\bf 55}, 6760 (1997); V.\ A.\ Kosteleck\'y, C.\ D.\ Lane, and A.\ G.\ M.\ Pickering, Phys.\ Rev.\ D\ {\bf 65}, 056006 (2002).

\bibitem{FWtransform} V.A.\ Kosteleck\'y and C.D.\ Lane, J.\ Math.\ Phys.\ {\bf 40}, 6245 (1999).

\bibitem{Bailey04} Q.G.\ Bailey and V.A.\ Kosteleck\'y, Phys.\ Rev.\ D {\bf 70}, 076006 (2004).

\bibitem{Parthey11} C.G.\ Parthey \etal, Phys.\ Rev.\ Lett.\ {\bf 107}, 203001 (2011).

\bibitem{Altschul10} B.\ Altschul, Phys.\ Rev.\ D {\bf 81}, 041701(R) (2010).

\end{thebibliography}
\end{document}